\documentclass[11pt,letterpaper]{article}
\pdfoutput=1
\usepackage[utf8]{inputenc}
\usepackage{amsmath}
\usepackage{amssymb}
\usepackage{graphicx}
\usepackage{multirow}
\usepackage{epsfig}
\usepackage{rotating}
\usepackage{longtable} 
\usepackage{url}
\usepackage{pdflscape}
\usepackage{cancel}
\pdfoutput=1
\usepackage{csquotes}
\usepackage{placeins}
\usepackage{scalerel}

\newcommand{\ie}{{\emph i.e.,\ }}

\newcommand{\yqu}{\mathcal{Y}_{q_L}^1}
\newcommand{\ylu}{\mathcal{Y}_{l_L}^1}
\newcommand{\yqt}{\mathcal{Y}_{q_L}^3}
\newcommand{\ylt}{\mathcal{Y}_{l_L}^3}
\newcommand{\yuu}{\mathcal{Y}_{u_R}^1}
\newcommand{\ydu}{\mathcal{Y}_{d_R}^1} 
\newcommand{\yut}{\mathcal{Y}_{u_R}^3}
\newcommand{\ydt}{\mathcal{Y}_{d_R}^3}
\newcommand{\ynu}{\mathcal{Y}_{\nu_R}^1}
\newcommand{\yeu}{\mathcal{Y}_{e_R}^1}
\newcommand{\ynt}{\mathcal{Y}_{\nu_R}^3}
\newcommand{\yet}{\mathcal{Y}_{e_R}^3} 
\newcommand{\yhu}{\mathcal{Y}_{\phi_1}}
\newcommand{\yhd}{\mathcal{Y}_{\phi_2}} 
\newcommand{\xl}[1]{\mathcal{Z}_{l_#1 }}
\newcommand{\xq}[1]{\mathcal{Z}_{q_#1 }}
\newcommand{\yu}{\mathcal{Y}}
\newcommand{\yd}{\mathcal{X}}
\newcommand{\gu}{g_{\mathcal{Y}}}
\newcommand{\gd}{g_{\mathcal{X}}}
\newcommand{\ysm}{Y_{SM}}
\newcommand{\ghatu}{\hat{g}_{\mathcal{Y}}}
\newcommand{\ghatd}{\hat{g}_{\mathcal{X}}}

\usepackage{amsthm}
\usepackage{xcolor}
\definecolor{nicered}{rgb}{0.7,0.1,0.1}
\definecolor{nicegreen}{rgb}{0.1,0.5,0.1}
\usepackage[colorlinks=true,citecolor= nicegreen,linkcolor=nicered]{hyperref}
\usepackage{cancel}

\usepackage{color}
\usepackage{soul}

\title{\large\sc {\bf
 Minimal non-universal  EW extensions of the Standard Model: a chiral  multi-parameter solution}
}

\author{
{\sc
{Richard H. Benavides}%
$^{a,}$\footnote{email: richardbenavides@itm.edu.co},~%
{Luis Muñoz}
$^{a,}$\footnote{email: luismunoz@itm.edu.co},~%
{William A. Ponce}
$^{b,}$} \footnote{email: william.ponce@udea.edu.co}~,\\%
{\sc
{Oscar Rodr\'iguez}%
$^{b,}$\footnote{email: oscara.rodriguez@udea.edu.co}~,%
and {Eduardo Rojas}%
$^{b,}$\footnote{email: rojas@gfif.udea.edu.co}%
}
\\
\\
{\sl $^{a}$ Facultad de Ciencias Exactas y Aplicadas, Instituto Tecnológico Metropolitano,}
 \\{\sl  Calle 73 No 76 A - 354 , Vía el Volador, Medellín, Colombia}\\
{\sl $^{b}$ Instituto  de F\'isica, Universidad de Antioquia,}\\ 
{\sl     Calle 70 No.~52-21, Medell\'in, Colombia}
}

\begin{document}
\maketitle

\begin{abstract}
We report the most general expression for the chiral 
charges of a non-universal $U(1)'$ with identical charges 
for the first two families but different charges  for the third one. 
The model is minimal in the sense that only standard model fermions 
plus  right-handed neutrinos are required. 
By imposing anomaly cancellation and constraints coming 
from Yukawa couplings we obtain two different solutions. 
In one of these solutions, the anomalies cancel between 
fermions in different families. 
These solutions depend  on four independent   parameters which
result very useful for model building. 
We build different  benchmark models in order to show the flexibility of the parameterization. 
We also report LHC and low energy constraints for these  benchmark models. 
\end{abstract}

\section{Introduction}
In the present work we address the question:  
what is the minimal electroweak extension of 
the standard model~(SM) with a minimal content of fermions? By itself,  this question is  interesting 
and deserves a dedicated and systematic study.
 The current literature on minimal models 
 abounds  in examples~\cite{Ponce:1987wb,He:1990pn,He:1991qd,Appelquist:2002mw,Carena:2004xs,Langacker:2008yv,Salvioni:2009jp,Crivellin:2015lwa,Ma:2016zod,Kownacki:2016pmx}
 but a general  parameterization of these models is not present in the literature, as far as we know.  
From a phenomenological point of view, 
owing to the absence of exotic fermions at low energies,
the minimal models are useful to explain  isolated anomalies  at  low energy experiments~(for a recent example of these kind of 
anomalies see~\cite{Pohl:2010zza,Aaij:2013qta,Krasznahorkay:2015iga,Heister:2016stz}). 

For universal models, that is, models in which  the hypercharge quantum  numbers are repeated for each  family, 
only  a trivial solution   with charges proportional 
to the SM hypercharge is possible if exotic fermions are not considered~\cite{Ponce:1987wb,Appelquist:2002mw,Carena:2004xs,Langacker:2008yv}.   
For non-universal models, as it is present  in the literature~\cite{He:1990pn,He:1991qd,Salvioni:2009jp,Crivellin:2015lwa}, 
the total number of parameters increases, given rise to  a large  variety of solutions.

The theoretical motivation  to study the 
non-universal models comes from  top-bottom approaches,
especially in string theory derived constructions, where  the $U(1)'$ charges
are family dependent~\cite{Langacker:2008yv}. 
Non-universal models have been also used  
to explain the number of families 
and the hierarchies in the fermion spectrum observed in the nature~\cite{Pisano:1991ee,Frampton:1992wt,Mantilla:2016lui}.

For gauge structures  with an  extended Electroweak~(EW)  sector~\cite{Langacker:2008yv},
the heavy vector bosons  $Z'$ associated with new  $U(1)'$ symmetries   are  generic  predictions of  physics Beyond the Standard Model~(BSM).
The detection of one of these resonances at the LHC  will shed light on
the underlying symmetries  of the BSM physics. 
For the high luminosity regime,
the  LHC will have  sensitivity for   $Z'$ masses below   5 TeV~\cite{Salazar:2015gxa,Godfrey:2013eta};
thus, a systematic and   exhaustive study  of the  EW extensions   
of the SM  with a minimal content of exotic ingredients  is convenient.  
By imposing  universality on the  EW  extensions of the SM (as it happens in the SM),  
the possible EW extensions are basically   $E_6$ subgroups~\cite{Carena:2004xs,Erler:2009jh,Erler:2011ud,Rodriguez:2016cgr}. 
It is well known  that  realistic scenarios for symmetry breaking in $E_6$  require 
large Higgs representations in order to explain the flavor phenomenology~\cite{Babu:2015psa}.
By relaxing  the universality constraints it is possible
 to have  small  Higgs and fermion  representations.
In this  case the anomaly cancellation can 
occur between fermions in different families; 
 among the most known  models  for three families are those  related to the local gauge 
structure $SU(3)_c\otimes SU(3)_L\otimes U(1)_x$  (3-3-1 for short)~\cite{Pisano:1991ee,Frampton:1992wt,Montero:1992jk,Foot:1992rh,Foot:1994ym,Ozer:1995xi,Ponce:2001jn,Ponce:2002sg,Okada:2015bxa,Cao:2016uur,Queiroz:2016gif}. 
For  flavor  models without electric exotic charges, \ie 
by restricting the values for the electric charges 
to those of the SM, the   classification of 3-3-1 models was presented in~\cite{Ponce:2001jn}. 
By allowing any rational value for the electric charge   an infinite number of models
is allowed, as it  was shown in~\cite{Diaz:2004fs,CarcamoHernandez:2005ka}.

Universality must not be taken for granted in models with physics beyond the SM. 
In particular, under some suitable assumptions
many non universal models are able to evade 
the Flavor Changing Neutral Currents~(FCNC) constraints.
In the present work we want to make a revision of  the different   $Z'$ models 
with a minimum content of fermions  and consistent with
the SM phenomenology;  owing  to the fact that these models are nonuniversal,
they  result very useful to explain some of  the recent 
flavor anomalies at the  LHCb~\cite{Aaij:2013qta,Aaij:2013iag,Jager:2014rwa}.

The paper is organized as follows: in Section~\ref{sec:2} we derive 
the general expressions for the   chiral charges of the models
for two different  scenarios, which correspond to  two 
different ways to cancel anomalies; 
in Section \ref{sec:3} we define  several 
benchmark models and it is pointed out which  coordinates  
in the parameter space  correspond to models previously studied  in the literature. 
  In Section \ref{sec:4} we derive the  95\% C.L. allowed 
 limits on the model parameters  by the most recent
 LHC data  and the corresponding limits by the  low energy electroweak data. 
  Section \ref{sec:5} summarizes our conclusions.
 

\section{The $SU(2)_L\otimes U(1)\otimes U(1)'$ gauge symmetry}
\label{sec:2}
The  aim  of the present work  is to build the most general parameterization
for the minimal electroweak extension of the  SM, limiting ourselves 
to the SM fermions plus right handed neutrinos.
In order to accomplish our purpose it is necessary
to give up universality; 
with this in mind, let us consider 
the gauge group $SU(2)\otimes U(1)\otimes U(1)'$ as a non-universal 
anomaly-free extension of the Electroweak sector of the Standard Model. 

In what follows $\mathcal{T}_{1L}$, $\mathcal{T}_{2L}$ and  $\mathcal{T}_{3L}$ denote the generators of $SU(2)_L$, while $\yu$ and $\yd$ denote the generators of $U(1)$ and $U(1)'$
respectively. For this gauge structure, the electric charge operator $Q$ must be a linear combination in the following way 
\begin{align}\label{Eq1}
Q=\mathcal{T}_{3L}+\dfrac{a}{2}\yu+\dfrac{b}{2}\yd,
\end{align}
where
\begin{align}\label{Eq2}
a\yu+b\yd=Y_{SM},
\end{align} 
being $Y_{SM}$ the hypercharge of the SM  and $a$ and $b$ are real parameters. Because  $Y_{SM}$ is known 
for every multiplet of the SM, and we have not assumed the existence of exotic particles, except the right handed neutrino, 
from the above equation we  can write $\mathcal{X}$ as a linear combination of $Y_{SM}$ and $\mathcal{Y}$, 
in such a way that the free parameters of the  model are  reduced to the  $\mathcal{Y}$ values
for the  SM Fermions, the right handed neutrinos and the  Higgs bosons.
 In what follows we can  avoid any  reference to  the specific values of  $\mathcal{X}$.
 The notation used for the   $\mathcal{Y}$ values  of the bosons and the fermions 
 of the first and third families are shown in Table~(\ref{tabla:1}).\\\\
The covariant derivative for our model is given by 
\begin{align}\label{Eq3}
D_{\mu} = \partial_{\mu}-ig\overrightarrow{\mathcal{T}}_L\cdot\overrightarrow{A}_{\mu}-i\dfrac{\gu}{2}\yu B_{\yu\mu}-i\dfrac{\gd}{2}\yd B_{\yd\mu},
\end{align} 
where  $g$, $\gu$ and $\gd$ are  the gauge  couplings asociated  to the  gauge groups $SU(2)_L$, $U(1)$ and $U(1)'$, respectively,
and  $\overrightarrow{A}_{\mu}$, $B_{\yu\mu}$ and $B_{\yd\mu}$ stand for the corresponding  gauge fields.

In order to avoid  the strong constraints coming from FCNC,
the first and second families have the same quantum numbers, but those of the third family are different (see Table~(\ref{tabla:1})). 
Because of this, at least two Higgs doublets are required in order  to give masses to the three families,
\begin{align}
\left\langle\phi_i\right\rangle^T=(0,v_i/\sqrt{2}),\hspace{0.5cm} i=1,2. 
\end{align}
In the next section we shall establish the necessary conditions to obtain an anomaly-free model. To this end we shall consider the fermion content of the SM extended with three right-handed neutrinos (one per family). 
\begin{table}
\begin{center}
\bgroup                    
\def\arraystretch{1.2}
\begin{tabular}{|c|c c c c c c c|}
\hline  
           &$\psi_L=(\nu, e)_L$  &$\nu_R$   &$e_R$   &$\psi '_L=(u, d)_L$  &$u_R$   &$d_R$    &$\phi_{1,2}$\\          
\hline
$Y_{SM}$   &$-1$       &$0$              &$-2$           &$1/3$       &$4/3$          &$-2/3$     &$1$  \\   
$\yu$      &$\yu^{1(3)}_{l_L}$ &$\yu^{1(3)}_{\nu_R}$  &$\yu^{1(3)}_{e_R}$ &$\yu^{1(3)}_{q_L}$  &$\yu^{1(3)}_{u_R}$  &$\yu^{1(3)}_{d_R}$ &$\yu_{\phi_{1,2}}$ \\  
\hline
\end{tabular}
\egroup
\caption{$U(1)$ charges for the chiral fields of the first (third) family and the two Higgs doublets. The charges for the second family are the same as those of the first one. SM hypercharges are also shown.} 
\label{tabla:1}
\end{center}
\end{table}
\subsection{Anomaly cancellation}
For the $SU(2)_L\otimes U(1)\otimes U(1)'$ symmetry the non-trivial anomaly equations are:
\begin{align}\label{Eq4}
[SU(2)]^2U(1):\hspace{0.2cm}2(&\yqu +\dfrac{1}{3}\ylu )+\yqt +\dfrac{1}{3}\ylt =0,\notag\\
[SU(3)]^2U(1):\hspace{0.2cm}2(&2\yqu -\yuu - \ydu)+2\yqt -\yut -\ydt =0,\notag\\
[\text{grav}]^2U(1):\hspace{0.2cm}2(&6\yqu -3\yuu -3\ydu+2\ylu -\ynu -\yeu )\notag\\
             +&6\yqt -3\yut -3\ydt +2\ylt -\ynt -\yet =0,\notag\\
[U(1)']^2U(1):\hspace{0.2cm}2(&\yqu -8\yuu -2\ydu +3\ylu -6\yeu )\notag\\
             +&\yqt -8\yut -2\ydt +3\ylt -6\yet =0,\notag\\
U(1)'[U(1)]^2:\hspace{0.2cm}2[&(\yqu)^2-2(\yuu)^2+(\ydu)^2-(\ylu)^2+(\yeu)^2]\notag\\
             +&(\yqt)^2-2(\yut)^2+(\ydt)^2-(\ylt)^2+(\yet)^2=0,\notag\\
[U(1)]^3:\hspace{0.2cm}2[&6(\yqu)^3-3(\yuu)^3-3(\ydu)^3+2(\ylu)^3-(\ynu)^3-(\yeu)^3]\notag\\
          +&6(\yqt)^3-3(\yut)^3-3(\ydt)^3+2(\ylt)^3-(\ynt)^3-(\yet)^3=0.
\end{align} 
From these equations and from  Eq.~(\ref{Eq2}), it can be shown that the other possible equations; that is, those corresponding to 
$[SU(2)]^2U(1)'$, $[SU(3)]^2U(1)'$, $[\text{grav}]^2U(1)'$ and $[U(1)']^3$ cancel out trivially.
We  also take into account the constraints coming from Yukawa couplings, 
\begin{align}
\mathcal{L}_Y\supset& \overline{l}_{1_L}\tilde{\phi}_1\nu_{1_R}+\overline{l}_{1L}\phi_1e_{1_R}+\overline{q}_{1_L}\tilde{\phi}_1u_{1_R}+\overline{q}_{1_L}\phi_1d_{1_R}+\notag\\
&\overline{l}_{3_L}\tilde{\phi}_2\nu_{3_R}+\overline{l}_{3_L}\phi_2e_{3_R}+\overline{q}_{3_L}\tilde{\phi}_2u_{3_R}+\overline{q}_{3_L}\phi_2d_{3_R}+\text{h.c}.
\end{align}
 The corresponding terms  of  the second family generate identical constraints as  those of the first family, for this reason they have 
 not been considered in the former equation. 
The corresponding  constraints  coming from  the  terms in the above Lagrangian are: 
\begin{align}\label{Eq5}
\yhu - \ynu + \ylu =0, \notag\\
\yhu + \yeu - \ylu =0, \notag\\
\yhu - \yuu + \yqu =0, \notag\\
\yhu + \ydu - \yqu =0, \notag\\
\yhd - \ynt + \ylt =0, \notag\\
\yhd + \yet - \ylt =0, \notag\\
\yhd - \yut + \yqt =0, \notag\\
\yhd + \ydt - \yqt =0. 
\end{align}
By solving simultaneously  the Eqs.~(\ref{Eq4}) and (\ref{Eq5}) we  find two  solutions (see Table~(\ref{tabla:0})). One of them corresponds to
what we  call scenario ${\bf A }$, in which the anomaly cancellation occurs in each family, while in  the another solution the anomaly cancellation takes place between 
fermions in different families; from now on, we will call this solution  scenario ${\bf B }$.  In both 
cases the $U(1)$  fermion charges   can be written in terms of four free-parameters, 
which we choose by convenience as 
$\left\lbrace \ynu, \ynt, \yqu, \yqt\right\rbrace$. 
As  a particular feature we observe that in scenario {\bf B} the $U(1)$ charges of the two Higgs-doublets turn out as a surprise  be equal,
for this reason, in this case only one doublet is necessary in  order to provide mass to the fermion fields, although  a singlet is   needed in  order 
to properly  break the gauge symmetry. 

As mentioned above,  to break $SU(2)\otimes U(1)\otimes U(1)'$ down to $U(1)_{Q}$, a minimal set of one $SU(2)$ doublet plus a singlet  is required. But
to properly generate viable quark masses  and a CKM mixing matrix, at least  a second doublet must be introduced. 
The generation of lepton~(neutrino) masses is more involved and may require new scalars, but it is a highly model dependent subject~\cite{Ma:1997nq}. 
However, there are two general cases of interest. The first one is the canonical type I seesaw where the
$\nu_R$ charges are set to zero. As we will see later, 
this condition is realized in the $Z_{\text{min}}$ model. 
An alternative  way  would be to forbid the  Dirac Yukawa coupling for the $\nu_R$. This would be
relevant to models in which a Dirac mass is generated by higher-dimensional operators and/or loops.
A detailed study of these extensions will be presented elsewhere.
\\

In the next section we  will calculate the  chiral couplings  of the SM fermions to the  $Z'$ boson.
\begin{table}
\begin{center}
\bgroup                    
\def\arraystretch{2.0}
\begin{tabular}{|c|c|c|}
\hline  
                &Scenario ${\bf A }$                            &Scenario ${\bf B }$ \\          
\hline
$\yhu$    &$3\yqu + \ynu$               &$2\yqu +\yqt + \dfrac{1}{3}(2\ynu + \ynt)$\\                
$\yhd$    &$3\yqt + \ynt$               &$2\yqu +\yqt + \dfrac{1}{3}(2\ynu + \ynt)$\\
$\ylu$         &$-3\yqu$                            &$-2\yqu - \yqt +\dfrac{1}{3}(\ynu - \ynt)$\\
$\yeu$     &$-6\yqu -\ynu$              &$-2(2\yqu +\yqt) -\dfrac{1}{3}(\ynu + 2\ynt)$\\
$\yuu$     &$4\yqu + \ynu$               &$3\yqu +\yqt +\dfrac{1}{3}(2\ynu + \ynt)$\\
$\ydu$     &$-2\yqu -\ynu$              &$-\yqu -\yqt - \dfrac{1}{3}(2\ynu + \ynt)$\\
$\ylt$         &$-3\yqt$                            &$-2\yqu -\yqt -\dfrac{2}{3}(\ynu - \ynt)$\\
$\yet$     &$-6\yqt -\ynt$              &$-2(2\yqu +\yqt) -\dfrac{1}{3}(4\ynu - \ynt)$\\
$\yut$     &$4\yqt + \ynt$               &$2(\yqu +\yqt) + \dfrac{1}{3}(2\ynu + \ynt)$\\
$\ydt$     &$-2\yqt - \ynt$              &$-2\yqu -\dfrac{1}{3}(2\ynu + \ynt)$\\
\hline
\end{tabular}
\egroup
\caption{Solutions to the  anomally cancellation equations  (\ref{Eq4}) and the  Yukawa constraints~(\ref{Eq5}).
The first solution (Scenario ${\bf A }$) corresponds to a framework where the anomaly cancellation occurs in each family separately.
For the another  solution (Scenario ${\bf B }$) the anomaly cancellation takes place between  fermions  in different families.
Notice that all the solutions are presented as functions of only the four parameters $\yqu$, $\yqt$, $\ynu$ and $\ynt$. } 
\label{tabla:0}
\end{center}
\end{table}

\subsection{Chiral charges}
The interaction between the fundamental fermions and the EW fields is given by the Lagrangian: 
\begin{align}\label{Eq8}
\mathcal{L}_{EW}= \sum_f i(&\overline{f}_{L} \gamma^\mu D_\mu f_{L} + \overline{f}_{R} \gamma^\mu D_\mu f_{R}), 
\end{align}
where $f$ runs over all fermions. By using  equation~(\ref{Eq3}) 
for the covariant derivative, 
and limiting ourselves  to those terms corresponding to the neutral gauge bosons,
the above expression can  then be   written as
\begin{align}\label{Eq9}
\mathcal{L}_{NC}= gJ^{\mu}_{3L} \mathcal{A}_{3\mu} + \gu J^{\mu}_\yu B_{\yu\mu} + \gd J^{\mu}_\yd B_{\yd\mu},
\end{align}
with
\begin{align}
J^{\mu}_\yu=&\dfrac{1}{2}\sum_f \overline{f}\gamma^{\mu}\left[\yu(f_L)P_L + \yu(f_R)P_R\right]f,\hspace{1cm}\text{and}\notag\\
J^{\mu}_\yd=&\dfrac{1}{2}\sum_f \overline{f}\gamma^{\mu}\left[\yd(f_L)P_L + \yd(f_R)P_R\right]f.
\end{align}
The values of $\yu$ for the different chiral states can be read off from Table~(\ref{tabla:1}),
and by using the relation~(\ref{Eq2})  it is possible to know  the corresponding values for $\yd$.\\\\
At this point we carry out an orthogonal 
transformation to write the original gauge fields $(B_\yu, B_\yd)$ in terms 
of the new gauge bosons $(B, Z')$, that is,
\begin{align}\label{Eq10}
B_{\yu\mu} &= \cos\theta B_{\mu} - \sin\theta Z'_{\mu}, \notag \\
B_{\yd\mu} &= \sin\theta B_{\mu} + \cos\theta Z'_{\mu},
\end{align}
being $\theta$ the mixing angle and $B_{\mu}$ the gauge field associated with
the SM hypercharge. In this new basis the neutral current Lagrangian Eq.~(\ref{Eq9}) is:
\begin{align}\label{Eq11}
\mathcal{L}_{NC} = gJ^{\mu}_{3L} \mathcal{A}_{3\mu} + g_{\ysm}J^{\mu}_{\ysm} B_{\mu} + g_{Z'}J^{\mu}_{Z'} Z'_{\mu},
\end{align}
where
\begin{align}\label{Eq12}
g_{\ysm}J^{\mu}_{\ysm} =&\hspace{0.4cm} \gu J^{\mu}_\yu\cos\theta+\gd J^{\mu}_\yd\sin\theta,\notag\\
 g_{Z'}J^{\mu}_{Z'}    =& -g_\yu J^{\mu}_\yu\sin\theta+g_\yd J^{\mu}_\yd\cos\theta, \notag\\
                       =&\hspace{0.5cm} g_{Z'}\sum_f \overline{f}\gamma^{\mu}\left[\epsilon_L(f)P_L + \epsilon_R(f)P_R\right]f.
\end{align}
In the last expression we have defined
\begin{align}\label{eq:chiral}
g_{Z'}\epsilon_L(f) &=\dfrac{1}{2}\left[-g_\yu\sin\theta \yu(f_L)+g_\yd\cos\theta \yd(f_L)\right],\notag\\
g_{Z'}\epsilon_R(f) &=\dfrac{1}{2}\left[-g_\yu\sin\theta \yu(f_R)+g_\yd\cos\theta \yd(f_R)\right].
\end{align}
Since Eq.~(\ref{Eq2}) implies the relation $aJ^{\mu}_\yu+bJ^{\mu}_\yd=J^{\mu}_{\ysm}$, 
the Eq.~(\ref{Eq12}) leads us to the following relations:
\begin{align}
ag_{\ysm}=& \hspace{0.1cm} g_\yu\cos\theta,\notag\\
bg_{\ysm}=& \hspace{0.1cm} g_\yd\sin\theta.
\end{align}
By defining $\ghatu\equiv \gu/a$ and $\ghatd\equiv \gd/b$, the above expressions are equivalent to
\begin{align}
\dfrac{\ghatu}{\ghatd}=&\tan\theta,\notag\\
\dfrac{1}{(g_{\ysm})^2}=&\dfrac{1}{(\ghatu)^2}+\dfrac{1}{(\ghatd)^2}.
\end{align} 
As can be shown by an explicit calculation, the chiral charges in Eq.~(\ref{eq:chiral}) can all be written as  linear combinations of the following four new  parameters
\begin{align}\label{Params}
\xl1 &\equiv \yu^1_{\nu_R}D,\notag\\
\xl3&\equiv \yu^3_{\nu_R}D,\notag\\
\xq1&\equiv C-3\yu^1_{q_L}D,\notag\\
\xq3&\equiv C-3\yu^3_{q_L}D,
\end{align}
where
\begin{align}
D=&\hspace{0.1cm}\dfrac{a(\ghatd)^2}{\sqrt{(\ghatd)^2-g^2_{\ysm}}}, \notag\\
C=& \sqrt{(\ghatd)^2-g^2_{\ysm}}.
\end{align}
By adopting these definitions in Table~(\ref{tabla:0}),   Eq.(~\ref{eq:chiral}) allowed us to obtain  the chiral charges  in  scenarios ${\bf A }$  and ${\bf B }$, which are shown in   Tables~(\ref{tabla:3})  and (\ref{tabla:4}), 
respectively.\\\\

\begin{table}
\begin{center}
\bgroup                    
\def\arraystretch{1.3}
\begin{tabular}{|c|c|c|c|}
\hline  
$f$           &$g_{Z'}\epsilon_L(f)$   &$g_{Z'}\epsilon_R(f)$                           &$g_{V}\epsilon^{V}_{L,R}$\\          
\hline
$\nu_{\alpha}$&$-\frac{1}{2}\xq\alpha$ &$-\frac{1}{2}\xl\alpha$                         &$-\frac{1}{2}\xl\alpha $ \\                
$e_{\alpha}$  &$-\frac{1}{2}\xq\alpha$ &$+\frac{1}{2}\left( \xl\alpha-2\xq\alpha\right)$&$-\frac{1}{2}\xl\alpha $ \\
$u_{\alpha}$  &$+\frac{1}{6}\xq\alpha$ &$-\frac{1}{6}\left(3\xl\alpha-4\xq\alpha\right)$&$+\frac{1}{6}\xl\alpha $ \\
$d_{\alpha}$  &$+\frac{1}{6}\xq\alpha$ &$+\frac{1}{6}\left(3\xl\alpha-2\xq\alpha\right)$&$+\frac{1}{6}\xl\alpha $ \\
\hline
\end{tabular}
\egroup
\caption{In the second and third  columns are shown the  chiral charges
which are obtained by requiring  anomaly cancellation  in each  family~(scenario ${\bf A }$).
By imposing that  the left chiral charges be equal to the 
right ones we obtain the most general  model with vector charges  in  scenario ${\bf A }$.  
 $\xl\alpha$  and  $\xq\alpha$ are arbitrary real parameters as can be seen in Eq.~(\ref{Params}).
 For $\xl1=\xl2=\xl3$ we obtain the universal $B-L$ model. $\alpha=1,2,3$ is a family index.} 
\label{tabla:3}
\end{center}
\end{table}

\begin{table}
\begin{center}
\setlength{\tabcolsep}{1pt}
\bgroup                    
\def\arraystretch{2}
\begin{tabular}{|c|c|c|}
\hline  
$f$          &$g_{Z'}\epsilon_L(f)$   &$g_{Z'}\epsilon_R(f)$\\        
\hline
$\nu_{1}$    &$-\frac{1}{6}\left(\xl1-\xl3+2\xq1+\xq3\right)$  &$-\frac{1}{2}\xl1$\\                
$e_{1}$      &$-\frac{1}{6}\left(\xl1-\xl3+2\xq1+\xq3\right)$  &$+\frac{1}{6}\left( \xl1+2\xl3-4\xq1-2\xq3\right)$\\    
$u_{1}$      &$+\frac{1}{6}\xq1$                               &$-\frac{1}{6}\left(2\xl1+ \xl3-3\xq1- \xq3\right)$\\
$d_{1}$      &$+\frac{1}{6}\xq1$                               &$+\frac{1}{6}\left(2\xl1+ \xl3-\xq1-  \xq3\right)$\\
\hline
$\nu_3$      &$+\frac{1}{6}\left(2\xl1-2\xl3-2\xq1-\xq3\right)$&$-\frac{1}{2}\xl3$\\                
$e_3$        &$+\frac{1}{6}\left(2\xl1-2\xl3-2\xq1-\xq3\right)$&$+\frac{1}{6}\left(4\xl1-\xl3-4\xq1-2\xq3\right)$\\
$u_3$        &$+\frac{1}{6}\xq3$                               &$-\frac{1}{6}\left(2\xl1+\xl3-2\xq1-2\xq3\right)$\\
$d_3$        &$+\frac{1}{6}\xq3$                               &$+\frac{1}{6}\left(2\xl1+ \xl3-2\xq1   \right)$\\
\hline
\end{tabular}
\egroup
\caption{In the second and third  columns are shown the  chiral charges
which are obtained by requiring  anomaly cancellation between fermions in different  families~(scenario ${\bf B }$).
 $\xl\alpha$  and  $\xq\alpha$ are arbitrary real parameters as can be seen in Eq.~(\ref{Params}).} 
\label{tabla:4}
\end{center}
\end{table}

\section{Benchmark models}
\label{sec:3}

\begin{table}
\begin{center}
\bgroup                    
\def\arraystretch{2.5}
\begin{tabular}{|c|l|l|}
\hline
Model                      & Definition                                            &       Constraints on $\xl\alpha$ and $\xq\alpha$ \\
\hline  
$Z_{V}^{{\bf A }}$            &\multirow{ 2}{*}{$\epsilon(f)_L=\epsilon_R(f)$}                         &$\xq\alpha = \xl\alpha$                         \\            
$Z_{V}^{{\bf B }}$            &                                                                        &$\xl3 =-2 \xl1 + 2 \xq1 + \xq3$                 \\ 
\hline 
$Z_{\tau}^{{\bf A }}$           &\multirow{ 2}{*}{$\epsilon_{L,R}(e_\beta)=\epsilon_{L,R}(\nu_\beta)=0$} &$\xl\beta=\xq\beta=0$    \\ 
$Z_{\tau}^{{\bf B }}$           &                                                                        &$\xl3 = 2 \xq1 + \xq3,\hspace{0.3cm} \xl1 = 0$    \\     
\hline
$Z_{\cancel{L}}^{{\bf B }}$     &$\epsilon_{L,R}(e_\alpha)=\epsilon_{L,R}(\nu_\alpha)=0$                 &$\xl1 = \xl3 = 0,\hspace{0.3cm} \xq3 = -2 \xq1$ \\
\hline
 $Z_{\cancel{p}}^{{\bf A }}  $  &\multirow{ 2}{*}{$2g_V(u)+g_V(d)=0$}                                    & $3\xq1 = \xl1$                                               \\
 $Z_{\cancel{p}}^{{\bf B }}  $  &                                                                        & $\xl3=-2 \xl1 + 8 \xq1 + \xq3$                 \\
\hline
 $Z_{\cancel{n}}^{{\bf A }}  $  &\multirow{ 2}{*}{$g_V(u)+2g_V(d)=0$}                                    & $\xq1 = -\xl1$                                                \\
 $Z_{\cancel{n}}^{{\bf B }}  $  &                                                                        & $\xl3=-2 \xl1 - 4 \xq1 + \xq3$                 \\ 
\hline                    
$Z_{t}^{{\bf B }}  $            & $\epsilon_{L,R}(u_\beta)=\epsilon_{L,R}(d_\beta)=0$                    &  $\xq1 = 0, \xq3 = 2 \xl1 + \xl3$                \\
\hline
 $Z_{\cancel{B}}^{{\bf B }}  $  &  $\epsilon_{L,R}(u_\alpha)=\epsilon_{L,R}(d_\alpha)=0$                 &  $\xq1 = 0,\hspace{0.3cm} \xq3 = 0,\hspace{0.3cm} \xl3 = -2 \xl1$          \\ 
\hline  
$Z_{\text{min}}^{({\bf A, B })}  $   & $\epsilon_{R}(\nu_\alpha)=0$                                        &  $\xl\alpha = 0  $    \\
\hline
\end{tabular}
\egroup
\caption{By imposing constraints on the chiral charges in Tables~(\ref{tabla:3}) and (\ref{tabla:4}) it is possible 
to define  benchmark models which result quite useful in the analysis of the experimental constraints.
The parameters $\xl\alpha$ and $\xq\alpha$ are arbitrary real numbers as can be seen  in Eq~(\ref{Params}).
 $\alpha=1,2,3$ and $\beta=1,2$.
In scenario ${\bf A}$  the charges of some benchmark models are equal to zero, for this reason, these possibilities are not shown here. } 
\label{tab:models}
\end{center}
\end{table}
The most general solution of the anomaly equations 
which satisfy  the constraints coming from the  Yukawa couplings 
depends on  four parameters. 
In general it is quite difficult to put constraints on this 
 4-dimensional space; however,  
it is possible to put very conservative constraints on 
some linear combinations of these parameters by using  benchmark models,
some of them already discussed in the literature. Let us see some examples~(All the models considered in this work are presented in Table~(\ref{tab:models})).

\begin{table}
\begin{center}
\bgroup                    
\def\arraystretch{1.3}
\begin{tabular}{|c|c|c|c|c|}
\hline  
$f$       &$g_{V}\epsilon^{V}_{L,R}$           &$g_{\tau}\epsilon^{\tau}_{L,R}$          &$g_{\cancel{L}}\epsilon^{\cancel{L}}_R$&$g_{t}\epsilon^{t}_{L,R}$  \\            
\hline
$\nu_1$   &$-\frac{1}{2}\xl1 $                 &$0$                                      &$0$                                    &$-\frac{1}{2}\xl1 $                      \\
$e_1$     &$-\frac{1}{2}\xl1 $                 &$0$                                      &$0$                                    &$-\frac{1}{2}\xl1 $                      \\
$u_1$     &$+\frac{1}{6}\xq1 $                 &$+\frac{1}{6}\xq1 $                      &$+\frac{1}{6}\xq1 $                    &$0$                                      \\
$d_1$     &$+\frac{1}{6}\xq1 $                 &$+\frac{1}{6}\xq1 $                      &$+\frac{1}{6}\xq1 $                    &$0$                                      \\
\hline
$\nu_3$   &$-\frac{1}{2}(2\xq1 +\xq3 -2\xl1 )$ &$-\frac{1}{2}(2\xq1 +\xq3 )$             &$0$                                    &$-\frac{1}{2}(\xq3 -2\xl1 )$             \\
$e_3$     &$-\frac{1}{2}(2\xq1 +\xq3 -2\xl1 )$ &$-\frac{1}{2}(2\xq1 +\xq3 )$             &$0$                                    &$-\frac{1}{2}(\xq3 -2\xl1 )$             \\
$u_3$     &$+\frac{1}{6}\xq3 $                 &$+\frac{1}{6}\xq3 $                      &$-\frac{1}{3}\xq1 $                    &$+\frac{1}{6}\xq3 $                      \\
$d_3$     &$+\frac{1}{6}\xq3 $                 &$+\frac{1}{6}\xq3 $                      &$-\frac{1}{3}\xq1 $                    &$+\frac{1}{6}\xq3 $                      \\
\hline
\end{tabular}
\egroup
  \caption{In the second column are shown the chiral charges for the
  most general   model  $Z_{V}$ with vector charges in scenario  ${\bf B }$,   
  from this model it is possible to get the chiral charges for the tauphilic $Z_{\tau}$, 
  leptophobic $Z_{\cancel{L}}$ and the  $Z_{t}$,  which are shown  in the third, 
  fourth and fifth columns, respectively. Here the charges 
  depend on three parameters $(\xl1 ,\xq1 ,\xq3 )$, which are defined in Eq.~(\ref{Params}).} 
\label{tab:bminuslb2}
\end{center}
\end{table}

\begin{table}
\begin{center}
\bgroup                    
\def\arraystretch{1.3}
\begin{tabular}{|c|l|l||l|l|}
\hline
          & \multicolumn{2}{|c||}{$Z_{\text{min}}^{{\bf B }}$ }  &  \multicolumn{2}{|c|}{$Z_{\cancel{p}}^{{\bf B }}$} \\
\hline  
$f$       &$g_{\text{min}}\epsilon^{\text{min}}_L$  &$g_{\text{min}}\epsilon^{\text{min}}_R$ &$g_{\cancel{p}}\epsilon^{\cancel{p}}_L$  &$g_{\cancel{p}}\epsilon^{\cancel{p}}_R$ \\            
\hline
$\nu_1$   &$-\frac{1}{6}(2\xq1+\xq3) $  &$0$                       &$-\frac{1}{2}(\xl1 -2\xq1 )$        &$-\frac{1}{2}\xl1 $                \\
$e_1$     &$-\frac{1}{6}(2\xq1+\xq3) $  &$-\frac{1}{3}(2\xq1+\xq3)$&$-\frac{1}{2}(\xl1 -2\xq1 )$        &$-\frac{1}{2}(\xl1 -4\xq1 )$       \\
$u_1$     &$+\frac{1}{6}\xq1 $          &$+\frac{ 1}{6}(3\xq1+\xq3)$&$+\frac{1}{6}\xq1 $                 &$-\frac{5}{6}\xq1 $                \\
$d_1$     &$+\frac{1}{6}\xq1 $          &$-\frac{1}{6}( \xq1+\xq3)$&$+\frac{1}{6}\xq1 $                 &$+\frac{7}{6}\xq1 $                \\
\hline
$\nu_3$   &$-\frac{1}{6}(2\xq1+\xq3)$   &$0          $             &$-\frac{1}{2}(6\xq1 +\xq3 -2\xl1 )$ &$-\frac{1}{2}(8\xq1 +\xq3 -2\xl1 )$\\
$e_3$     &$-\frac{1}{6}(2\xq1+\xq3)$   &$-\frac{1}{3}(2\xq1+\xq3)$&$-\frac{1}{2}(6\xq1 +\xq3 -2\xl1 )$ &$-\frac{1}{2}(4\xq1 +\xq3 -2\xl1 )$\\
$u_3$     &$+\frac{1}{6}\xq3 $          &$+\frac{1}{3}(\xq1+\xq3)$ &$+\frac{1}{6}\xq3                 $ &$-\frac{1}{6}(6\xq1 -\xq3 )$       \\
$d_3$     &$+\frac{1}{6}\xq3 $          &$-\frac{1}{3}\xq1$        &$+\frac{1}{6}\xq3 $                 &$+\frac{1}{6}(6\xq1 +\xq3 )$       \\
\hline
\end{tabular}
\egroup
\caption{In the second column, chiral charges for the  minimal  model $Z_{\text{min}}^{{\bf B }}$ are presented,
and in the third column are  the corresponding charges for the protonphobic model $Z_{\cancel{p}}^{{\bf B }}$.
In both cases the models belong to   scenario ${\bf B }$. 
Here the charges depend only  on three real arbitrary parameters $(\xl1 ,\xl3 ,\xq3 )$.} 
\label{tab:proton}
\end{center}
\end{table}

In order to cross-check our equations, it is convenient to calculate the charges for the most general  
$Z'$ model with vector charges $Z_{V}^{\bf A,B}$, in our framework these charges
are shown in Tables~(\ref{tabla:3}) and (\ref{tab:bminuslb2}) for the  scenarios  ${\bf A }$ and  ${\bf B }$  respectively. 
By using these charges  it is possible to reproduce 
the $Z_{B-L}$  model by  taking $\xl1=\xl3$ in  scenario ${\bf A }$,  and $\xl1=\xq1=\xq3$ in  scenario ${\bf B }$. 
The $Z_{B-L}$ model is the minimal universal  model with right-handed neutrinos with a vector-like neutral current. 
Another  model with  a vector-like  neutral current  is   the tau-philic model $Z_{\tau}$ which have zero couplings to the
leptons of the first and the second families, and  non-zero couplings  for  the $\tau$.  In  Tables~(\ref{tabla:3}) and 
(\ref{tab:bminuslb2}) this condition is met by setting 
$\xl1=0$. In this family, the model $B-3L_{\tau}$ is 
the best-known example in the  literature~\cite{Ma:1997nq,Ma:1998dp,Ma:1998dr}.
Modulo a global normalization, the charges of the $Z_{\tau}$ reduce to those of  $Z_{B-3L_{\tau}}$ 
by requiring   $\xq1=\xq3$ in Table~(\ref{tab:bminuslb2}).
This model was proposed to have radiative masses with acceptable phenomenological 
values for neutrino oscillations,  by allowing an extended scalar sector~~\cite{Ma:1997nq}.  
In reference~\cite{Pal:2003ip} was pointed out 
 that if there is a gauged $B-3L_{\tau}$    symmetry at low energy, 
 it can prevent fast proton decay. This model is also  able to provide  
 dark matter candidates as has been studied in~\cite{Okada:2012sp}.  
For the   scenario ${\bf A }$  a  chiral tau-philic model is also possible in a trivial way by making in Table~(\ref{tabla:3}) $\xq\alpha=\xl\alpha=0$  for the first 
and the second families (\ie for $\alpha=1,2$)  and  $\xq3\neq 0$  and $\xl3\neq 0$. 
Other  interesting family of  models is the   $Z_{t}$ which is defined to have  zero couplings to the quarks of the
first and second families but couplings  different from zero  for the  top and the bottom quarks.
An special subset of  models  in $Z_t$   are  the hadrophobic models $Z_{\cancel{B}}$ which have zero
couplings to the quarks of the three families. 
Indeed, $Z'$ hadrophobic models attracted a lot of interest in connection with the $e^{\pm}$ excess in cosmic
ray data  observed by ATIC and  PAMELA  experiments~\cite{Salvioni:2009jp,Cirelli:2008pk,Fox:2008kb,Bi:2009uj}.
Another interesting model is the $Z_{\text{min}}^{({\bf A,B})}$ which has zero couplings to the right handed neutrinos, allowing a Majorana mass term.\\

\noindent
For dark matter interacting with the SM fermions through a $Z'$,
an  isospin violating interaction constitutes
a possible solution to some challenges posed by some experimental 
results~\cite{Kurylov:2003ra,Giuliani:2005my,Chang:2010yk,Kang:2010mh,Yaguna:2016bga}.
A maximal isospin violation is possible by requiring zero couplings  
to the proton but different from zero for the neutron or in the other way around. 
For a nucleus with $Z$ protons and $N$ neutrons the weak charge is given by
\begin{align}
Q_{W}(N,Z)=  Q_{W}(p) Z +Q_{W}(n)N,
\end{align}
Where $Q_{W}(p)=-2 (2 C_{1u}+C_{1d})$ and $Q_{W}(n)=-2 (2C_{1d} + C_{1u})$ are the proton and neutron weak charges, respectively. 
Here (for the definitions see references~\cite{Langacker:1991pg,Erler:2011iw,Rojas:2015tqa}) 
\begin{align}
 C_{1q} =2g_{A}^{(1)}(e)g_{V}^{(1)}(q)+ 2\left(\frac{g^{\prime}M_{Z}}{g^{(1)}M_{Z'}}\right)^2g_{A}^{\prime}(e)g_{V}^{\prime}(q) ,  \notag\\
 C_{2q} =2g_{V}^{(1)}(e)g_{A}^{(1)}(q)+ 2\left(\frac{g^{\prime}M_{Z}}{g^{(1)}M_{Z'}}\right)^2g_{V}^{\prime}(e)g_{A}^{\prime}(q),  
\end{align}
where  $g_{V,A}^{(1)}(f)$ and $g^{(1)}$ are the vector (axial) coupling  and the coupling strength, respectively,
of the fermion $f$ to the SM $Z$ boson and 
$g_{V,A}^{\prime}(f)$ and $g'$ are  the corresponding quantities for the interaction with  the $Z'$. 
The shift in the proton and neutron   weak charges owing to the $Z'$ couplings to the standard model fermions is  
\begin{align}
\Delta Q_{W}(p)= -4\left(\frac{g'M_{Z}}{g^{(1)}M_{Z'}}\right)^2 g_{A}^{\prime}(e)\left(2 g_{V}^{\prime}(u)+g_{V}^{\prime}(d)\right),\notag\\
\Delta Q_{W}(n)= -4\left(\frac{g'M_{Z}}{g^{(1)}M_{Z'}}\right)^2 g_{A}^{\prime}(e)\left(2 g_{V}^{\prime}(d)+g_{V}^{\prime}(u)\right). 
\end{align}
By requiring that $\Delta Q_{W}(p)=0$ (with $g_{A}(e)\ne 0$) we obtain the  
protonphobic model~\footnote{Our definitions of protonphobic 
and neutronphobic refer to
bosons which do not couple - at vanishing momentum transfer and at the tree level - to 
protons and neutrons, respectively. This definition is different from  the  definition  presented in reference~\cite{Feng:2016jff}.}  
$Z_{\cancel{p}}^{\bf A,B}$. 
The chiral charges for this model are shown in Table~(\ref{tab:proton}).
In an  identical  way we proceed to obtain the corresponding  charges of the   neutronphobic model
 $Z_{\cancel{n}}^{\bf A,B}$.

\section{LHC and low energy constraints}
\label{sec:4}
\begin{figure*}
\begin{center}
\centering 
\begin{tabular}{cc}
 \includegraphics[scale=0.3]{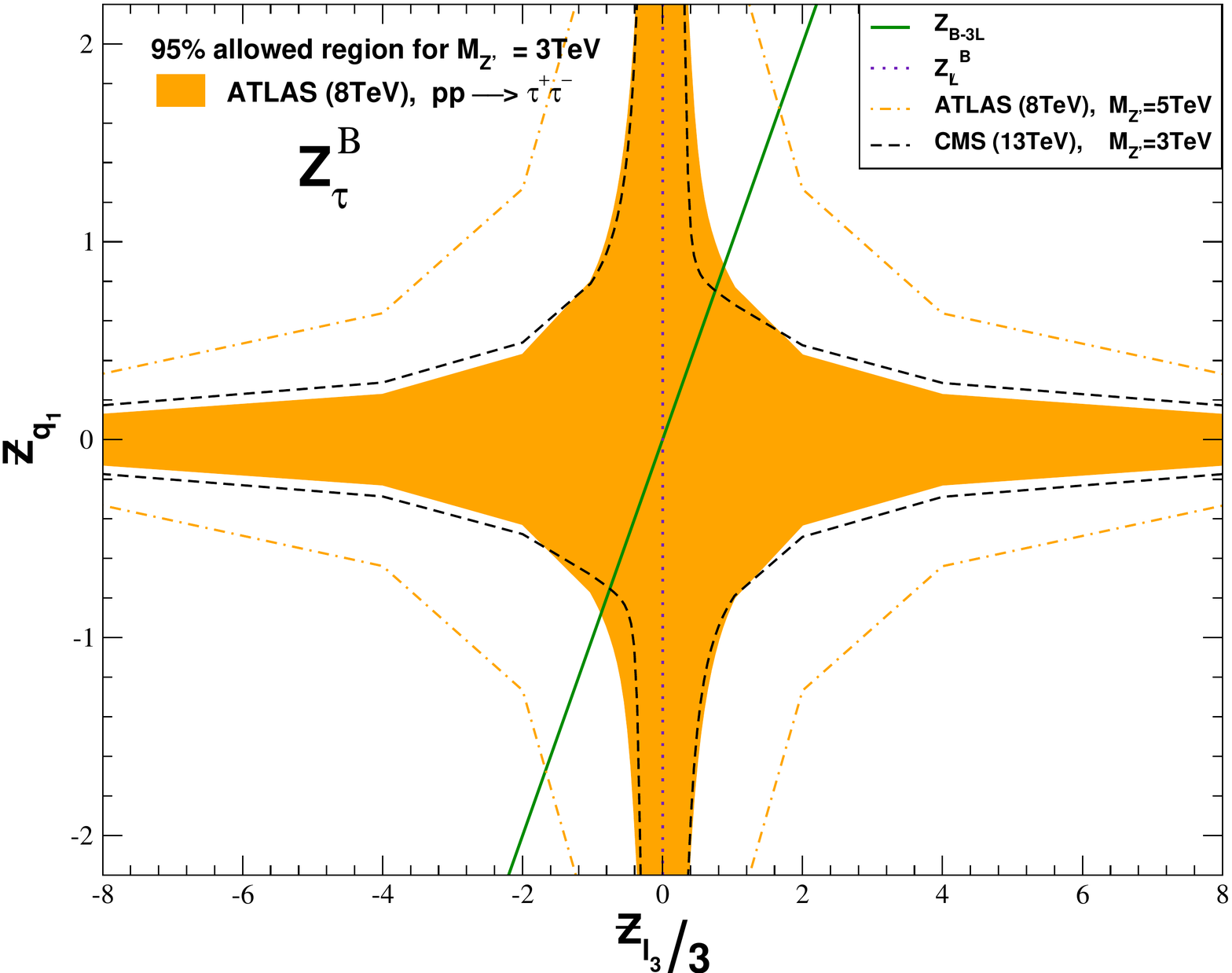}   &  \includegraphics[scale=0.3]{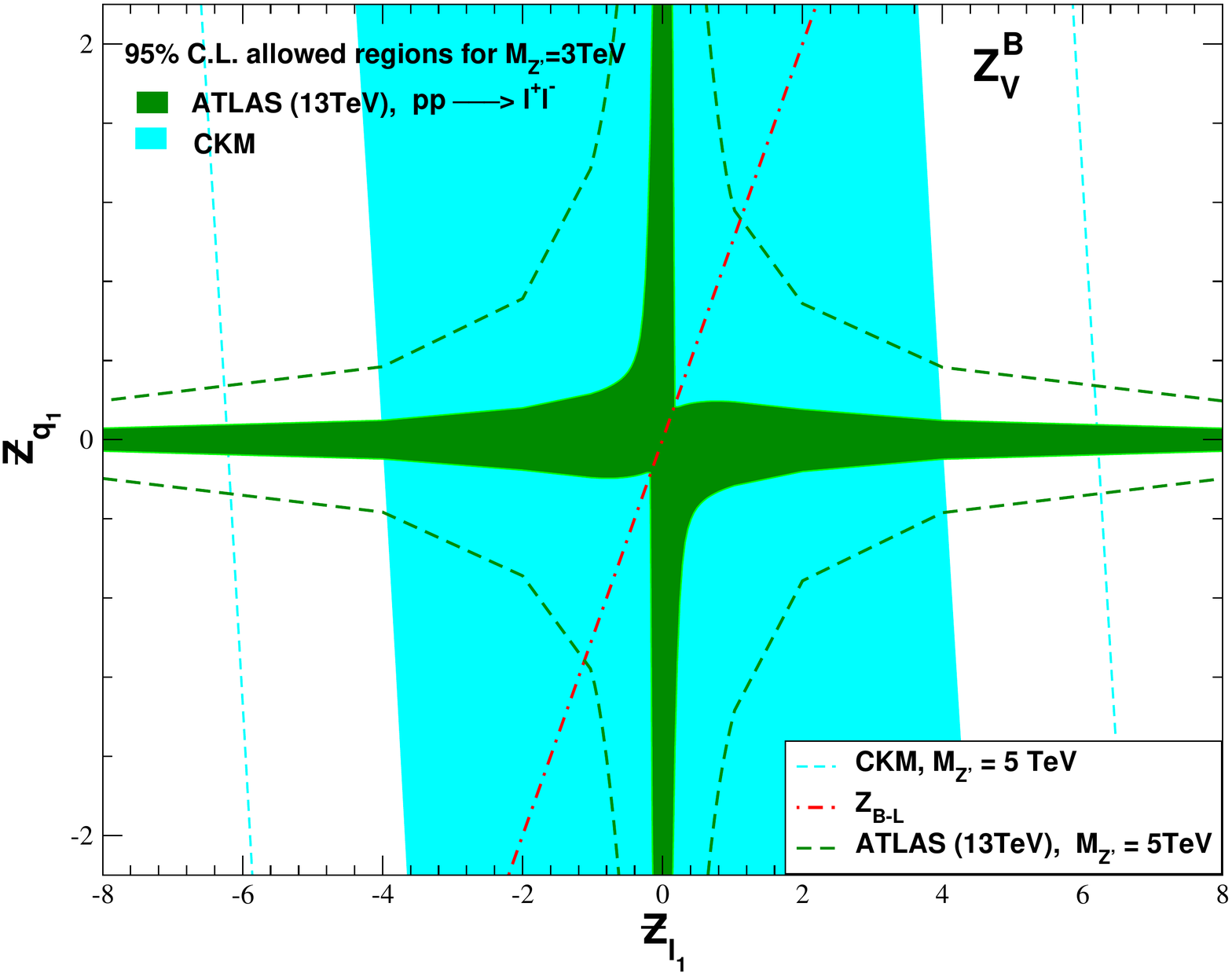}\\   
 \includegraphics[scale=0.31]{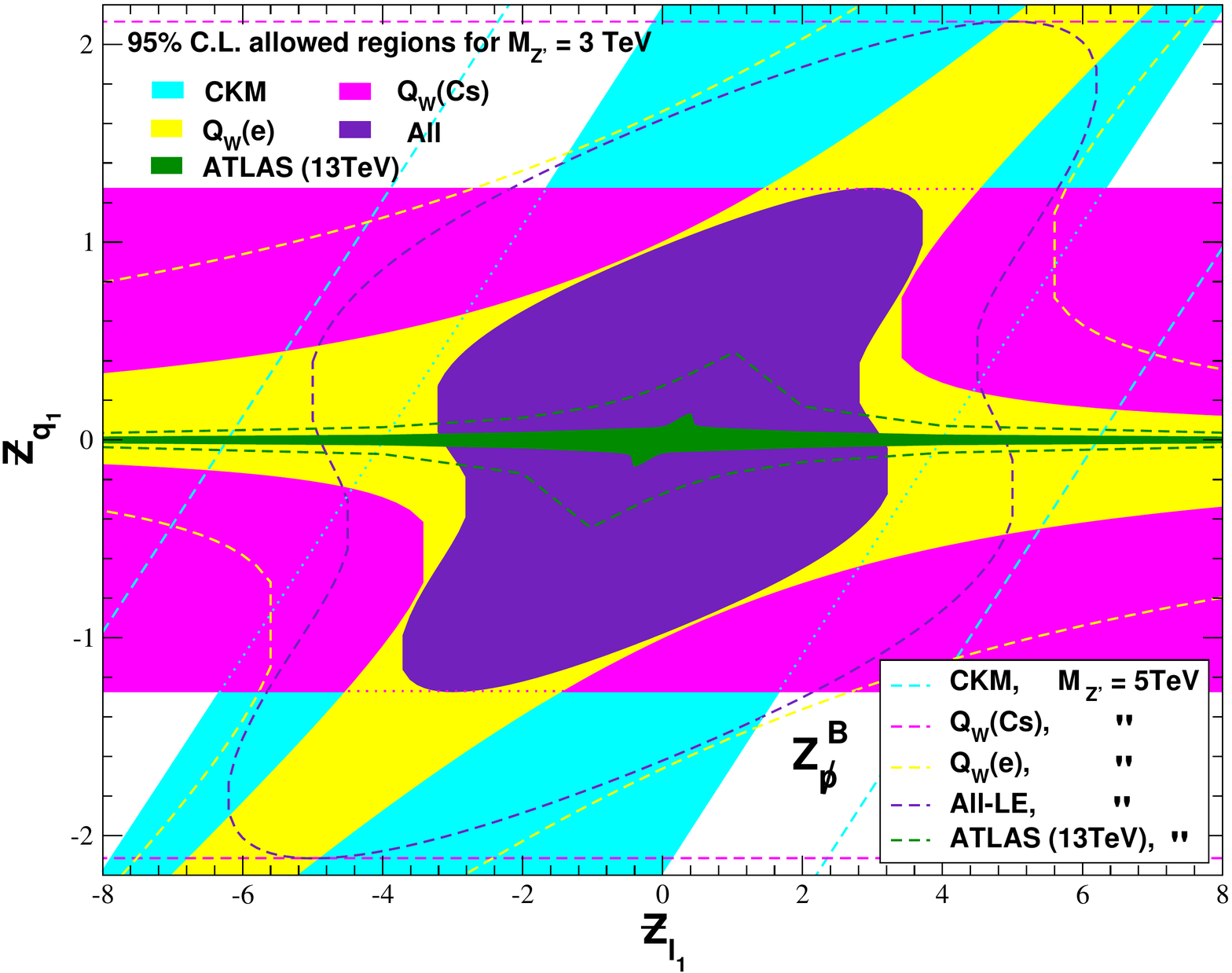} 
&\includegraphics[scale=0.31]{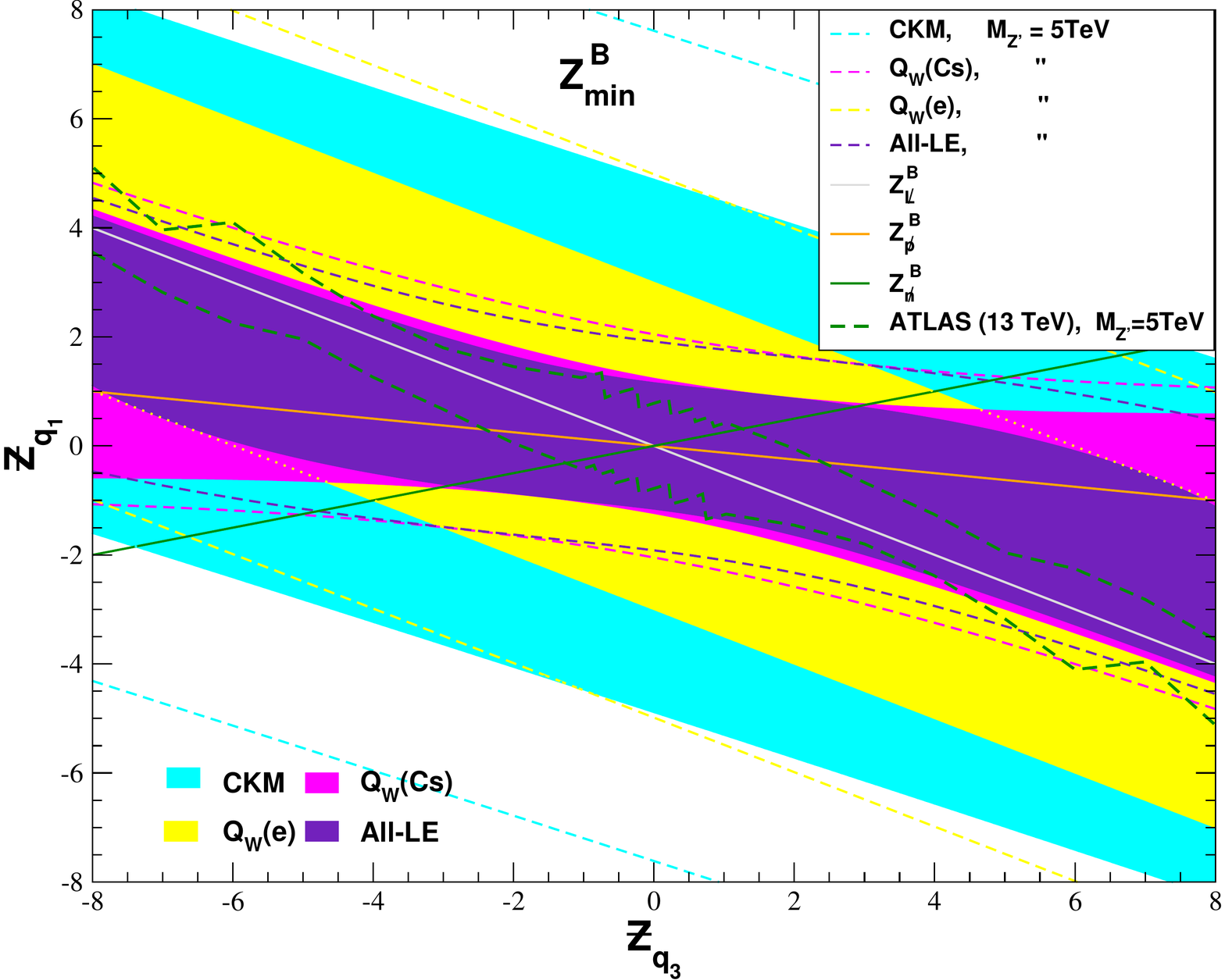}
\end{tabular}
\end{center}
\caption{
Colored regions correspond to  the allowed parameter space  at the 95\% C.L for a $M_{Z'}=3$TeV.   
The orange region  in  the left plot in the top panel corresponds to the 95\% C.L.  allowed by data from  
proton-proton collisions decaying to tau pairs in the ATLAS detector with an integrated luminosity of $19.5-20.3$~fb$^{-1}$. 
Contours are also shown for the same channel at 13TeV with a luminosity of 2.2~fb$^{-1}$ from CMS data. 
In the remaining plots the green region corresponds to the 95\% C.L. allowed region by data proton-proton collisions
decaying to electrons and dimuons with an integrated luminosity of  13.3~fb$^{-1}$,
the magenta region corresponds to the 95\% C.L. allowed region by 
 the  electron weak charge measurements in Moller scattering. The yellow region corresponds to the 95\% C.L.  allowed region 
by the cesium weak charge measurements.  The cyan region corresponds 
to the allowed region by the constraints on the violation of the first-row CKM unitarity~\cite{Agashe:2014kda}.
By combining all the data the 95\% C.L. allowed  parameter space   corresponds to the   indigo region. 
The region inside  the dashed magenta, yellow, cyan and indigo correspond to the 95\% allowed regions for a $M_{Z'}=5$TeV. 
The $Z_{\text{min}}^{\bf B}$ model is basically excluded for a $M_{Z'}=$3TeV, for this reason this contour is not shown. 
}
\label{Contours3}	
\end{figure*}

In this section we report  the most recent constraints, 
from colliders and low energy experiments, on the $Z'$    parameters  for  some  benchmark models.
For the time being, the strongest constraints  come  from the   proton-proton collisions  data, collected by the ATLAS experiment 
at the LHC  with an integrated luminosity of 13.3~fb$^{-1}$  
at a  center of mass energy of 13 TeV.  In particular, 
we used the upper limits  at 95\% C.L.  on the total cross-section of the $Z'$ decaying
into dileptons~\cite{ATLAS:2016cyf} (\ie  $e^+e^-$  and  $\mu^+\mu^-$).  In Figure~(\ref{Contours3}) the colored 
green regions correspond to the allowed regions for this data.

Even though the dilepton data   put the strongest constraints on 
three of the four models in Figure~(\ref{Contours3}) this data do not put 
limits on the parameters of the tauphilic model $Z_{\tau}$ 
because this model have zero couplings to the electron and the muon. 
For this model we used instead, the strongest constraints on 
the total cross-section $pp\rightarrow \tau^+\tau^- $ channel, 
which come from the  proton-proton collisions data, collected by the ATLAS experiment,   
at a center of mass energy of 8~TeV and an   integrated luminosity of 19.5-20.3~fb$^{-1}$~\cite{Aad:2015osa}.
For this channel the most recent constraints, with a  similar strength than 
those of ATLAS, come from the data collected by the CMS experiment  at a center 
of mass energy of 13~TeV and an integrated luminosity of 2.2~fb$^{-1}$~\cite{CMS:2016zxk,GonzalezHernandez:2016fir}..  
In figure~(\ref{Contours3}) the 95\% C.L.  allowed regions  by  the ATLAS and CMS data, 
for  the tauphilic parameters   are  shown.   

There is also possible to put constraints by using data  from low energy experiments. 
The low energy strongest constraints  come from Atomic Parity Violation~(APV), in particular from  the cesium weak charge~\cite{Wood:1997zq,Guena:2004sq} and 
the electron weak charge measurement by the SLAC-E158 collaboration~\cite{Anthony:2005pm}.
The experimental values and the analytical expressions for these observables are shown in Table~(\ref{tab:weakq}).
The APV observables  depend on the electron  axial coupling  
to the $Z'$ boson which is zero 
in the vector model $Z_{V}$ in consequence, there are not APV  limits on this model in  Figure~(\ref{Contours3}).  
An important constraint on  $Z_{V}$ comes from the limits   on the violation of the first-row CKM unitarity~\cite{Marciano:1987ja,Buras:2013dea}.
For this model the constraints on the  $\xq1$ parameter are dominated by the $pp\rightarrow l^+l^-$ channel; however,
this channel do not put  limits on the  $\xl1$ parameter for small values of $\xq1$; as can be seen in Figure~(\ref{Contours3})
in this case the CKM unitarity is able to put bounds even for $\xq1=0$. 
This plot shows  the importance of the low energy constraints in order to narrow 
the new physics parameters.

In order  to show the complementarity of some experiments 
the  constraints on  the parameter space for the protonphobic and neutronphobic models 
are shown  in Figure~(\ref{Contours3}). 
 \begin{table}
 \begin{center}
 \bgroup                    
 \def\arraystretch{1.3}
 \begin{tabular}{|c|c|c|c|}
 \hline  
 $\mathcal{Q}$                    &Value~\cite{Agashe:2014kda}                & SM prediction~\cite{Agashe:2014kda}       & $\Delta \mathcal{Q}$\\ 
  \hline
 $Q_W(\text{Cs})$                 &$-72.62\pm 0.43$     &$-73.25\pm 0.02$     & $Z\Delta Q_W(p)+N\Delta Q_W(n)$ \\
 \hline
 $Q_W(e)$                         &$-0.0403\pm 0.0053$  &$-0.0473\pm 0.0003$  & $-4\left(\frac{g' M_{Z}}{g^{(1)}M_{Z'}}\right)^2 g_A^{\prime}(e)g_{V}^{\prime}(e)$  \\
 \hline 
$1-\sum_{q=d,s,b}|V_{uq}|^2$      &$1-0.9999(6)$        &         0           & $\Delta_0\epsilon_L(\mu)\left(\epsilon_L(\mu)-\epsilon_L(d)\right)    $                                                                  \\          
 \hline
 \end{tabular}
 \egroup
 \caption{Experimental value and SM prediction of the Cesium and electron weak charges 
 and the respective shift owed to the interaction with  the $Z'$. 
  The third observable is the constraint on the violation of the first-row CKM unitarity~\cite{Agashe:2014kda} where
 $\Delta_0 =\frac{3}{4\pi^2}\frac{M_W^2}{M_{Z'}^2}\ln\frac{M_{Z'}^2}{M_W^2}g^{\prime 2}$.} 
 \label{tab:weakq}	
 \end{center}
 \end{table}
For some models, the low-energy observables can constrain one of the parameters in Eq.~(\ref{Params}) independently of the values of the remaining ones. These results are shown in Table~(\ref{tab:one-par}).
\begin{table}
\begin{center}
\bgroup                    
\def\arraystretch{2.0}
\begin{tabular}{|c|c|c|}
\hline  
Model                &$M_{Z'}$ = 3 TeV                            &$M_{Z'}$ = 5 TeV \\          
\hline
$Z^{\bf A}_{V}$              &$\vert \xl1\vert \leq 3.112$             &$\vert \xl1\vert \leq 4.856$\\                
$Z^{\bf A}_{\cancel{p}}$     &$\vert \xl1\vert \leq 3.558$             &$\vert \xl1\vert \leq 5.927$\\
$Z^{\bf A}_{\cancel{n}}$     &$\vert \xl1\vert \leq 0.856$            &$\vert \xl1\vert \leq  1.426$\\
$Z^{\bf A}_{\text{min}}$     &$\vert \xq1\vert \leq 1.180$              &$\vert \xq1\vert\leq 1.964$\\
$Z^{\bf B}_{t}$              &$\vert \xl1\vert \leq 3.594$             &$\vert \xl1\vert \leq 5.607$\\
$Z^{\bf B}_{\cancel{B}}$     &$\vert \xl1\vert \leq 3.594$             &$\vert \xl1\vert \leq 5.607$\\
\hline
\end{tabular}
\egroup
\caption{Bounds on models for which the low-energy observables can constrain one of the  parameters in Eq.~(\ref{Params}) independently of the values of the remaining ones.} 
\label{tab:one-par}
\end{center}
\end{table}

\section{Conclusions}
\label{sec:5}
In the present work, we presented the most general  chiral charges 
of  the minimal universal and non-universal  
$Z'$ model  with a minimal content of fermions.
Even though, several minimal models have been reported before, 
the complete solution as a function of a set of  continuous parameters
and its  corresponding collider and low energy  constraints, as far as we know,  
is a new result in the literature. 

In general, minimal  models are of  a great interest for the beyond  SM 
phenomenology~\cite{He:1990pn,He:1991qd,Salvioni:2009jp,Crivellin:2015lwa,Celis:2015ara,Almeida:2004hj,Demir:2005ti,Accomando:2016sge,Accomando:2016eom,Basso:2013vla}.
In particle physics the  Non-universal models  are well motivated,
especially in String theory derived constructions, 
where the $U(1)^{\prime}$ charges are family non-universal~\cite{Langacker:2008yv}.
Non-universal models have also  been  used 
to explain the number of families and 
the hierarchies in the fermion spectrum in the SM~\cite{Pisano:1991ee,Frampton:1992wt}. 
In our analysis we rule out some possibilities on phenomenological grounds limiting ourselves
to a couple of scenarios  to cancel the anomalies. 
 In the simplest case    
 or scenario {\bf A} the anomalies 
 cancel between fermions in every family.
 It is fairly obvious that  from this scenario, 
 it is possible to obtain, as a particular case, the charges of the minimal universal models
  which, as it is well known~\cite{Langacker:2008yv},  can be written as a linear combination of the charges of the  $Z_{B-L}$ model and the 
 SM hypercharge.

 In the second case or scenario {\bf B} the anomalies cancel
 between fermions in different families. 
 Although it is true that some particular  models in this scenario have been reported before, 
 to the best of our knowledge the full parameterization for this scenario is a new result in the literature. 
   To prevent FCNC constraints 
the charges of the first and second familues  were  assumed to be identical,
but different  to the charges of the third family.
Constraints from  the SM Yukawa interactions were used to impose 
additional constraints   in such  a way that the number of free parameters 
associated with the chiral charges was  reduced to four parameters.
We also report the most recent LHC constraints on
the parameter space for some benchmark models and 
compare them to those coming from experiments at low energies. 
From our analysis, we showed that the unitarity constraints on
the CKM are able to exclude some regions in the parameter space 
which are difficult to exclude by using only LHC data. 
\section*{Acknowledgments} 
R. H. B. and L. M. thank to \enquote{Centro de Investigaciones  ITM}. 
E.R.  thanks  C. Salazar for  technical assistance. 
We thank financial support from 
\enquote{Patrimonio Autónomo Fondo Nacional de Financiamiento para la Ciencia, la Tecnología
y la Innovación, Francisco José de Caldas}, 
and \enquote{Sostenibilidad-UDEA 2016}.

\FloatBarrier
\bibliographystyle{apsrev4-1longdoi}
\bibliography{references331}

\end{document}